\documentclass[prl,aps,superscriptaddress,twocolumn,showpacs,nofootinbib]{revtex4-1}
\usepackage{graphicx,color}
\usepackage{amsmath,amsfonts,enumerate,amsthm,amssymb,bbm}
\usepackage{hyperref}
\usepackage{multirow}
\usepackage{bbold}
\usepackage{braket}
\usepackage{soul}
\usepackage[caption = false]{subfig}
\usepackage{ulem}
\usepackage{scrextend}

\mathchardef\ordinarycolon\mathcode`\:
\mathcode`\:=\string"8000
\begingroup \catcode`\:=\active
  \gdef:{\mathrel{\mathop\ordinarycolon}}
\endgroup

\hypersetup{
   colorlinks=true,         
   linkcolor=blue,          
   citecolor=blue,          
}

\theoremstyle{plain}

\theoremstyle{definition}

\theoremstyle{remark}

\def\<{\langle}

\def\tr{ \mbox{tr} \,}
\def\>{\rangle}
\def\<{\langle}

\renewcommand{\v}[1]{\ensuremath{\boldsymbol #1}}

\newcommand{\be}{\begin{equation}}
\newcommand{\ee}{\end{equation}}

\newcommand{\new}[1]{{\color[RGB]{0,0,0}{#1}}}
\newcommand{\zo}[1]{{\color[RGB]{0,0,0}{#1}}}

\newcommand{\change}[1]{{\color[RGB]{0,0,0}{#1}}}

\newcommand{\drop}[1]{$\hdots$}

\begin{document}

\title{Enhanced energy transfer to an optomechanical piston from indistinguishable photons}

\author{Zo\"e Holmes}
\email{zholmes@lanl.gov}
\affiliation{Controlled Quantum Dynamics Theory Group, Imperial College London, Prince Consort Road, London SW7 2BW, UK.}
\affiliation{Physics and Astronomy, University of Exeter, Exeter, EX4 4QL, UK.}
\affiliation{Information Sciences, Los Alamos National Laboratory, Los Alamos, NM, USA.}
\author{Janet Anders}
\affiliation{Physics and Astronomy, University of Exeter, Exeter, EX4 4QL, UK.}
\affiliation{Institut f\"ur Physik und Astronomie, Potsdam University, 14476 Potsdam,  Germany.}
\author{Florian Mintert}
\affiliation{Controlled Quantum Dynamics Theory Group, Imperial College London, Prince Consort Road, London SW7 2BW, UK.}

\begin{abstract}

\new{\zo{Thought experiments involving gases and pistons, such as Maxwell's demon and Gibbs' mixing, are central to our understanding of thermodynamics. Here we present a quantum thermodynamic thought experiment in which the energy transfer from two photonic gases to a piston membrane grows quadratically with the number of photons for indistinguishable gases, while linearly for distinguishable gases.
This signature of Bosonic bunching may be observed in optomechanical experiments, highlighting the potential of these systems for the realization of thermodynamic thought experiments in the quantum realm.}}
\end{abstract}

\maketitle

\new{The concept of particle indistinguishability is deeply entwined into the history of both quantum mechanics and thermodynamics. The first remarkable example of the consequences of the difference between distinguishable and indistinguishable particles is found in Gibbs' thought experiment~\cite{Gibbs} on the extraction of work from the mixing of gases.
Subsequently, the indistinguishability of energy quanta played a central role in the development of quantum mechanics through Planck's reconciliation of Wien's law and the Rayleigh-Jeans limit of black body radiation.
The indistinguishability of elementary particles, \new{Fermions and Bosons, }is now recognised as a fundamental principle, with diverse signatures such as the Pauli blockade~\cite{PauliBlockade} or
the Hong-Ou-Mandel effect~\cite{HOM}, which causes even non-interacting photons to leave beamsplitters in pairs, {\it i.e.} to bunch.}

\new{The role of the statistics of indistinguishable quantum particles in thermodynamics has recently gathered renewed attention. For quantum generalisations of a Szilard engine the extractable work is independent of whether the working substance is Bosonic or Fermionic~\cite{SzilardStats}, but
Bosonic bunching can enhance the conversion of information and work~\cite{BunchingSzilardPRL} and the performance of thermodynamic cycles~\cite{PrasannaBunching, DeffnerBoson}.}

 \new{Although any two Fermions or Bosons of the same type are intrinsically identical, in practise it is often possible to distinguish such particles via their internal states~\cite{DiuCohenLaloe}.
In the case of photons, for example, the distinguishability can be carried by a degree of freedom such as polarization that admits coherent superpositions. The distinguishability between two photons, one vertically polarised and the other in the state $\ket{\theta}=\cos\theta\ket{V}+\sin\theta\ket{H}$ with $\ket{V}$ and $\ket{H}$ referring to vertical and horizontal polarizations, can thus be varied continuously, with the photons \textit{partially} distinguishable for $ 0 < \theta < \pi/2$.} 

\new{The possibility of partially distinguishable quantum gases has provided a natural generalisation to Gibbs mixing~\cite{Schrodinger,thesoviets} with many implications for thermodynamics.
For example, it is impossible to perfectly distinguish non-orthogonal quantum states without breaking the second law of thermodynamics~\cite{Peres},
the accessible information in Gibbs mixing is limited by the Holevo bound~\cite{Maruyama}, and the extractable work from mixing, defined as the \textit{ergotropy}~\cite{ergotropy}, can decrease with distinguishability~\cite{allahverdyan}.}

\zo{In this paper we present a thought-experiment that probes the interplay of distinguishability and particles statistics in quantum thermodynamics. 
Drawing inspiration from ground breaking thought experiments involving gases performing work on a membrane attached to a movable piston~\cite{Gibbs, Maxwell, Szilard}, we consider the interaction between photon gases and a beamsplitter membrane, that gives the photons access to superpositions of spatial states located either side of the membrane. We find that the photonic bunching results in striking consequences for how energy is transferred between light and the membrane. Namely, as a result of the Hong-Ou-Mandel effect, the energy transfer grows quadratically with the number of photons for indistinguishable gases, while linearly for distinguishable gases.}

\begin{figure}[t]
{\includegraphics[width =0.4\textwidth]{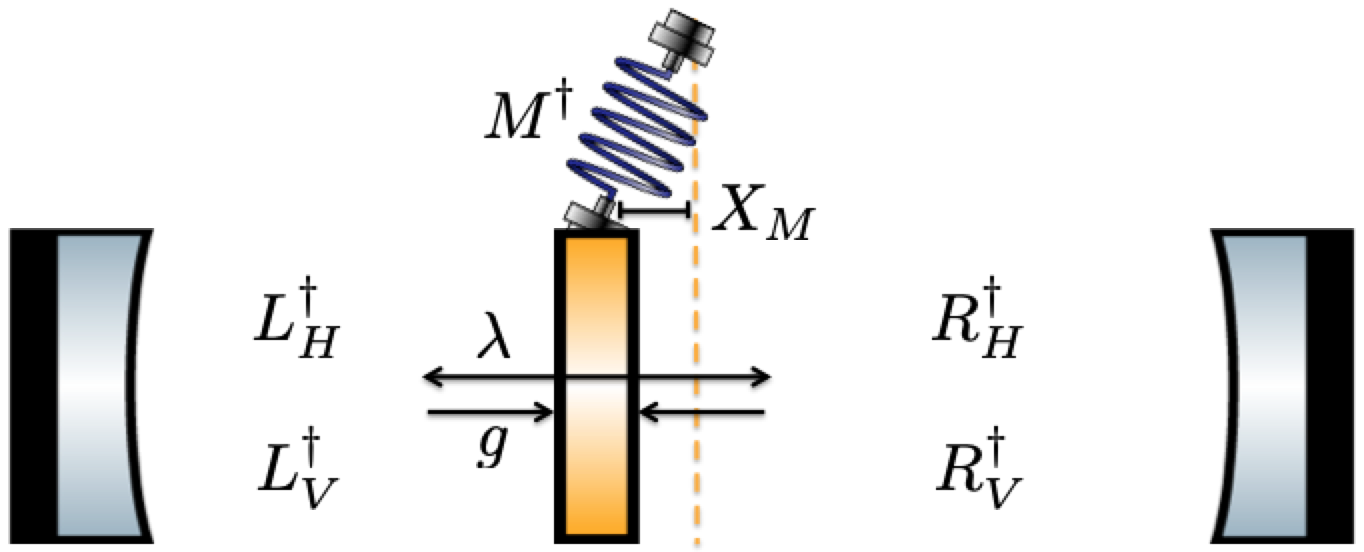}}
\caption{Illustration of an optomechanics setup to generalise a variety of classic thermodynamic experiments involving gases and piston membranes. The classical gases are realised by photon gases on either side of a movable beamsplitter membrane.}
\label{fig:optomechplot} 
\end{figure}  

\zo{The proposed thought experiment may be realised in multi-mode optomechanical systems in which a microscopic membrane separates an optical cavity into two parts~\cite{optomechbook, optomechreview, experiment1,experiment5,experiment2,experiment4}, thus highlighting a new avenue for quantum thermodynamic experiments. We argue here that such multi-mode optomechanical setups go beyond previous proposals in single-mode systems~\cite{Elouard_2015,OptoMechExperiment,OptoMechthermo4,OptoMechthermo3}, by providing a platform both for studying quantum signatures of distinguishability and, more broadly, realising thermodynamic thought experiments involving the interaction of gases with membranes.}

\medskip

A setup realising this variety of quantum mechanical generalizations of piston-like experiments is given by an optomechanical system comprised of a cavity with a membrane that behaves like a beamsplitter and that divides the cavity symmetrically into a left and a right part, as sketched in Fig.~\ref{fig:optomechplot}. 
The photon dynamics resultant from the membrane can be modelled in terms of the Hamiltonian
\begin{equation}
H_{\mbox{\tiny BS}} =\sum_{p=H,V}
\frac{\lambda}{2} \left(R_p^\dagger L_p + L_p^\dagger R_p\right)\ ,
\label{eq:bs}
\end{equation}
\new{where $\lambda$ is the intercavity coupling strength and the annihilation and the annihilation (creation) operators of both the horizontally, $R_{H}^{(\dagger)}$ and $L_{H}^{(\dagger)}$, and vertically, $R_{V}^{(\dagger)}$ and $L_{V}^{(\dagger)}$, polarised photons in the right and left halves of the cavity are explicitly modelled in order to study the effect of distinguishability~\cite{zeilinger,polarisationOptomech}}.
The membrane has a motional degree of freedom (DOF), like a cantilever,
and the interaction between the light field and the motional DOF is given by
\begin{equation}
H_{\mbox{\tiny I}} = - g (N_L-N_R) X_M\ ,
\label{eq:om}
\end{equation}
in terms of the total particle number in the left and right part of the cavity
({\it i.e.} $N_{L}=L_H^\dagger L_H + L_V^\dagger L_V$, $N_{R}=R_H^\dagger R_H + R_V^\dagger R_V$)
and the displacement operator $X_M$ of the membrane~\cite{experiment1a}. 
This optomechanical coupling will allow us to discuss a notion of energy (be it work or heat) transferred to the mechanical DOF in analogy to the extraction of work in the classical setting.

The full system Hamiltonian $H$ is given by the sum of $H_{\mbox{\tiny BS}}$ and $H_I$ (defined in Eqs.~\eqref{eq:bs} and \eqref{eq:om}) and the non-interacting terms for the four photonic modes $H_C = \omega(N_L+N_R)$ and single phonon mode $H_M =\omega_M M^\dagger M$. The eigenfrequencies of both parts of the cavity and of the mechanical DOF are denoted by $\omega$ and $\omega_M$ respectively and the annihilation (creation) operator $M^{(\dagger)}$ of the mechanical phonons is related to the displacement operator via 
$X_M= x_{\mbox{\tiny zpf}}( M+M^\dagger)$. The prefactor $x_{\mbox{\tiny zpf}}$ is the mechanical oscillator's zero point uncertainty $x_{\mbox{\tiny zpf}} = 1/\sqrt{2 m \omega_M}$ with $m$ the mass of the membrane.

To solve the system dynamics explicitly, despite the high-dimensional Hilbert space,
it is helpful to consider the equations of motion for the observables of interest in the Heisenberg picture.
The equation of motion for the displacement $X_M$ of the mechanical DOF,
\begin{align}\label{eq:eomX}
    \frac{d^2 X_M}{d t^2} +  2 \kappa_M \frac{d X_M}{d t} + \omega_M^2 X_M  =\frac{ g}{m} \left( \Delta N_{H}+\Delta N_{V} \right) \, ,
\end{align}
depends on the photonic mode imbalances
$\Delta N_p=L_p^\dagger L_p - R_p^\dagger R_p$ (for $p=H$ and $V$)
whose dynamics result from the equation of motion
\begin{align}
    &\frac{d L_p}{dt} = -i (\omega + g X_M - i \kappa) L_p - i \frac{\lambda}{2} R_p \, , \label{eq:eomL} \\ 
    &\frac{d R_p}{dt} = -i (\omega - g X_M - i \kappa) R_p - i \frac{\lambda}{2} L_p \; . \label{eq:eomR}
\end{align}
\new{In a thought experiment the damping rates $\kappa$ and $\kappa_M$ of the cavity and mechanical modes can be assumed to vanish such that the number of photons in each gas is conserved and the piston membrane is frictionless. However, any experiment would be realised with finite damping.}


Solving the coupled differential equations Eqs.~(\ref{eq:eomX}-\ref{eq:eomR}),
exactly is prohibitively difficult but a perturbative solution that describes the dynamics of selected observables for any given initial state can be constructed. \new{Given that the force exerted by a single photon on the membrane is weak, the single photon coupling strength, $g x_{\mbox{\tiny zpf}}$, is small compared to the inter-cavity coupling strength, $\lambda$. It is therefore appropriate to solve the dynamics perturbatively in $g$. \zo{With the membrane cooled to cryogenic temperatures (mean phonon occupation number $\bar{n}_{th}$ of the order of 10)~\cite{CooledMembrane}, the higher order contributions are negligible, as discussed in the supplement~\cite{supMat}. We thus focus here on the dynamics to first order; however, similar behaviour is observed to higher orders~\cite{supMat}}.}



The initial state of the left and right parts of the cavity and the membrane, $\rho_L \otimes \rho_R \otimes \sigma_M$, is chosen in analogy to classical thermodynamic thought experiments and hence the gases are taken to have the same number distribution (and therefore the same average photon number $\langle N(0) \rangle = \tr_L[ N_L (0) \, \rho_L ]= \tr_R[ N_R (0) \, \rho_R ]$ and variance $\delta N (0)$). The transition between distinguishable and indistinguishable photon gases can be explored by taking all photons in the left cavity to be in the polarisation state $\ket{V}$ and all photons in the right in $\ket{\theta}$, where $\theta$ can be varied continuously between $0$ to $\pi/2$. 
For the membrane, an initial state with vanishing displacement, $\tr_M[X_M(0) \, \sigma_M] =0$, and vanishing momentum is inline with the classical thought experiments, for example, the membrane could be prepared in a thermal state.

\zo{The dynamics induced by $H_{\mbox{\tiny BS}}$ entangles the mechanical and optical degrees of freedom, resulting in an energy transfer from the effective energy of the photons, $ (\omega - g X_M) N_L$ and  $ (\omega + g X_M) N_R$, to the membrane.}
To first order in the interaction constant $g$, the quantum mechanical average of the energy of the membrane is given by
\begin{equation}
\Delta H_M(t) = u(t) \, \delta N(0)  +  v(t) \left(\langle N(0) \rangle+ \langle N(0) \rangle^2 \cos^2(\theta) \right)\ ,
\label{eq:DH}
\end{equation}
following Eqs.~(\ref{eq:eomX}-\ref{eq:eomR}).
\zo{The scalar prefactors $u(t)$ and $v(t)$, discussed further in \cite{supMat}, are positive oscillatory functions (see Fig.~\ref{fig:EnergyTransfer}) that depend on the system parameters ($g, \omega_M, \lambda, m, \kappa$ and $\kappa_M$), but not on the initial state of the gases, which enters through the terms $\delta N(0)$, $\langle N(0) \rangle$ and $\cos^2(\theta)$.}

For any choice in the number distribution of the photon gases, the energy transfer to the membrane is larger for indistinguishable photons than distinguishable photons, with the difference between these two cases scaling as $\langle N(0) \rangle^2 \cos^2(\theta)$. That is, the energy transfer to the membrane is quadratically enhanced for indistinguishable photons. The enhancement is most pronounced for Fock states where the initial fluctuations in photon number $\delta N(0)$ vanish, or coherent states where the fluctuations are equal to the average photon number, $\delta N(0) = \langle N(0) \rangle$. Conversely, for high temperature thermal gases\footnote{\new{Here, and elsewhere when discussing thermal photon gases, we refer to a gas with a fixed polarisation but a thermally distributed photon number distribution.}} the initial fluctuations in photon number $\delta N(0)$ will be substantial and so there is a substantial contribution to $\Delta H_M$ that is independent of $\theta$. 

This dependence of the energy transfer on distinguishability is the opposite to Gibbs mixing where work extraction is possible for distinguishable gases but not for indistinguishable gases.
The difference in behaviour is perhaps unsurprising as the present mechanism does not rely on mixing.
What seems striking is the scaling with particle number. Whereas the extractable work in the Gibbs~\cite{Gibbs}, and indeed the Szilard~\cite{Szilard} and Maxwell Demon thought-experiments~\cite{Maxwell}, scales linearly with the particle number, {\it i.e.} it can be interpreted as `work per particle', the present situation realises a quadratic scaling, with a potentially strongly enhanced energy transfer to the membrane.

\begin{figure}[]
\includegraphics[width=\linewidth]{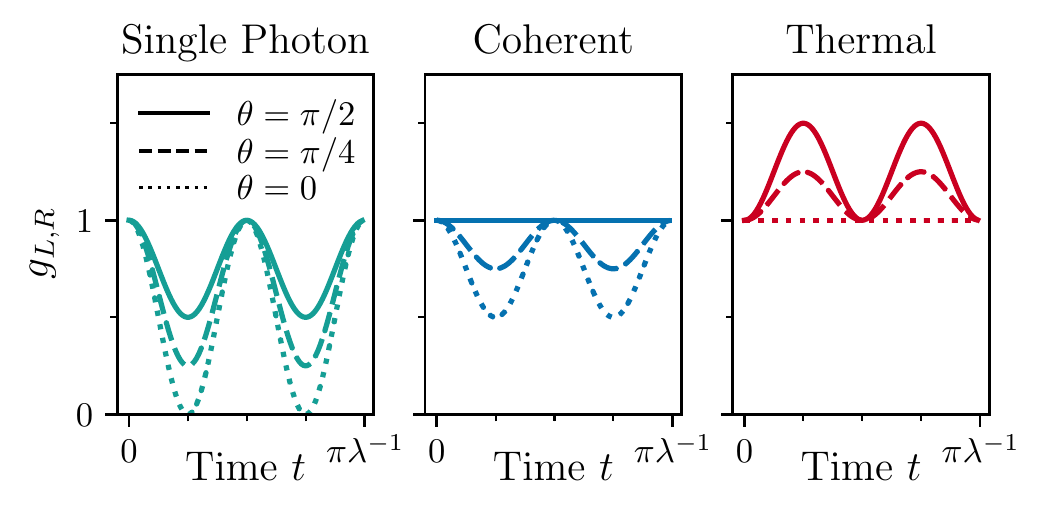}
\caption{\new{The correlation function $\mathbb{g}_{\mbox{\tiny L,R}}$, Eq.~\eqref{eq:g2}, as function of time for single photon, coherent and thermal initial states of the light field} \new{in the absence of damping ($\kappa = \kappa_M = 0$)}. Perfectly distinguishable ($\theta = \pi/2$), partially distinguishable ($\theta = \pi/4$) and perfectly indistinguishable ($\theta = 0$) gases are denoted by solid, dashed and dotted lines.}
\label{fig:g2}
\end{figure}

As we will show in the following, this quantum mechanical enhancement of energy transfer, $\Delta H_M$, between light and the mechanical DOF is a direct consequence of photon bunching as observed in the Hong-Ou-Mandel (HOM) effect~\cite{HOM,SaharHOM}.
To this end, it is instructive to inspect the \textit{two mode} second order correlation
function~\cite{SuppMattCorrelationFunc} 
\begin{equation}
\mathbb{g}_{\mbox{\tiny L,R}}(t)= \frac{\langle N_L(t)N_R(t)\rangle}{\langle N_L(t)\rangle  \langle N_R(t)\rangle}\ .
\label{eq:g2}
\end{equation}
A vanishing value of $\mathbb{g}_{\mbox{\tiny L,R}}$ indicates that a measurement would find all photons in one cavity, whereas large values of $\mathbb{g}_{\mbox{\tiny L,R}}$ imply that approximately equal numbers would be found in both halves of the cavity.
A small value of $\mathbb{g}_{\mbox{\tiny L,R}}$ thus indicates bunching, whereas a large value indicates anti-bunching~\cite{supMat}.

The dynamics of $\mathbb{g}_{\mbox{\tiny L,R}}$ as the light field interacts with the beamsplitter membrane can readily be obtained to first order in $g$.
It is depicted in Fig.~\ref{fig:g2} \new{in the absence of damping effects ($\kappa = \kappa_M = 0$)} for perfectly distinguishable ($\theta=\pi/2$), perfectly indistinguishable ($\theta=0$) and partially distinguishable ($\theta=\pi/4$) gases.
In all \new{three sub-figures, corresponding to single photon, coherent and thermal states of the light field,} one can see that for all times, distinguishable gases result in the largest values of $\mathbb{g}_{\mbox{\tiny L,R}}$ and indistinguishable gases the smallest.
Moreover, the time averaged correlation function~\cite{supMat},
\begin{equation}
   \langle \mathbb{g}_{\mbox{\tiny L,R}}(t) \rangle_t = \frac{1}{4} ( \gamma + 3 - \cos^2( \theta) ) \, ,
\end{equation}
where $\gamma = 2$ for thermal photons, $\gamma = 1 - \frac{1}{n}$ for an $n$ photon Fock state \new{and $\gamma = 1$ for coherent state photons}, has the same $\cos^2(\theta)$ dependence on distinguishability as the energy transfer to the membrane, Eq.~\eqref{eq:DH}.
That is, bunching is most pronounced for indistinguishable gases, as expected. 

\begin{figure}[t]
\includegraphics[width=\linewidth]{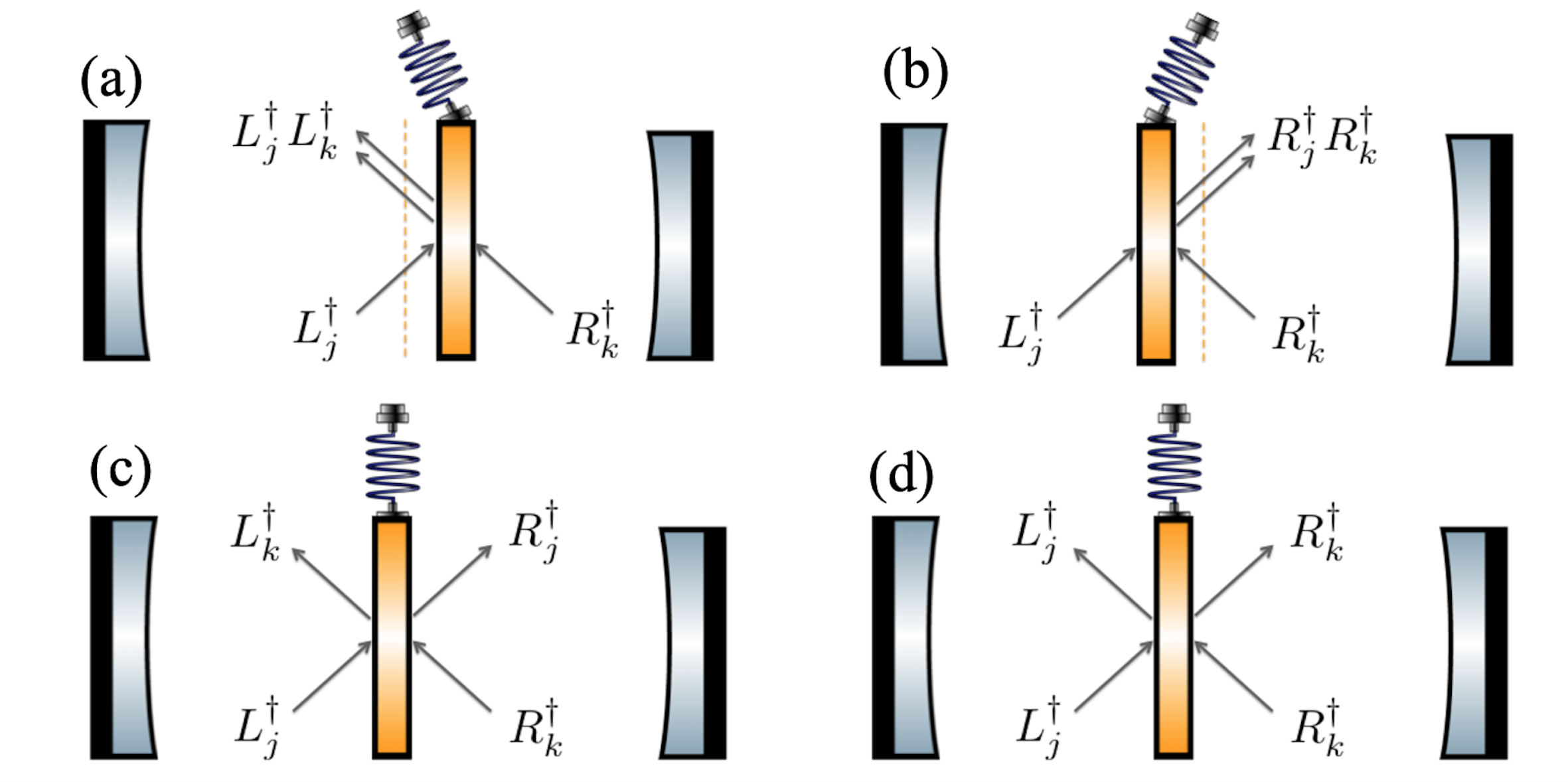}
\caption{\label{fig:HOMinCavity}
The Hong-Ou-Mandel effect in the present optomechanical setup.
 In (a) the photon from the left is reflected and the photon from the right is transmitted and vice versa in (b). In (c) both photons are transmitted and in (d) both are reflected. When the photons are perfectly distinguishable, i.e. $j = H$ and $k = V$ or vice versa, then all four outcomes (a - d) are equally probable. When the photons are perfectly indistinguishable, {\it i.e.} $j = H$ and $k = H$ or $j = V$ and $k = V$, then the amplitudes for outcomes (c) and (d) destructively interfere and the outcomes (a) and (b) are equally probable.}
\end{figure}

To understand heuristically how this bunching affects the membrane dynamics, it is instructive to consider
the case of single photon gases
as sketched in Fig.~\ref{fig:HOMinCavity}.
For both distinguishable and indistinguishable photons the (quantum) average displacement of the membrane will be zero \new{at all times}. However, the fluctuations in the position of the membrane\new{, and therefore the energy of the membrane,} will be greater for the case of indistinguishable photons because the probability for the membrane to be displaced to the left or right is double that for distinguishable photons. \new{Moreover, the quadratic scaling of the energy transfer may be explained by the fact that the HOM effect is a pairwise interference effect. Since $\langle N(0) \rangle$ photons in one gas can interfere with each of the $\langle N(0) \rangle$ photons in the other gas, the number of pairs of photons which can interfere with one another scales as \change{$\langle N(0) \rangle^2$} and this quadratic scaling carries over to the energy transfer.}

\new{
\new{While considering an initial thermal state for the photon gases realises a close analogy with classical thermodynamics,
including the process of pumping brings the discussion closer to an experimentally realizable situation.
If the cavity is driven on resonance in a pulsed fashion~\cite{PulsedVannerProp,PulsedVannerImp,PulsedNJP,Clarke_2018,PulsedClarke},
with pulses that are} shorter than the tunnelling time $\lambda$, the driving processes and tunnelling processes occur on different time scales and can be considered independently. Accordingly, driving the left modes of the cavity with a short laser pulse of $\theta$ polarised photons and the right modes with a short pulse of vertically polarised photons, will generate the coherent states $\ket{\alpha, \theta}$ and $\ket{\alpha, V}$ in the respective halves of the cavity~\cite{Fox}, leading again to an enhanced energy transfer to the membrane for indistinguishable photons as per Eq.~\eqref{eq:DH} with $\delta N = \langle N \rangle = |\alpha|^2$.

\begin{figure}[t]
\centering
\includegraphics[width=\linewidth]{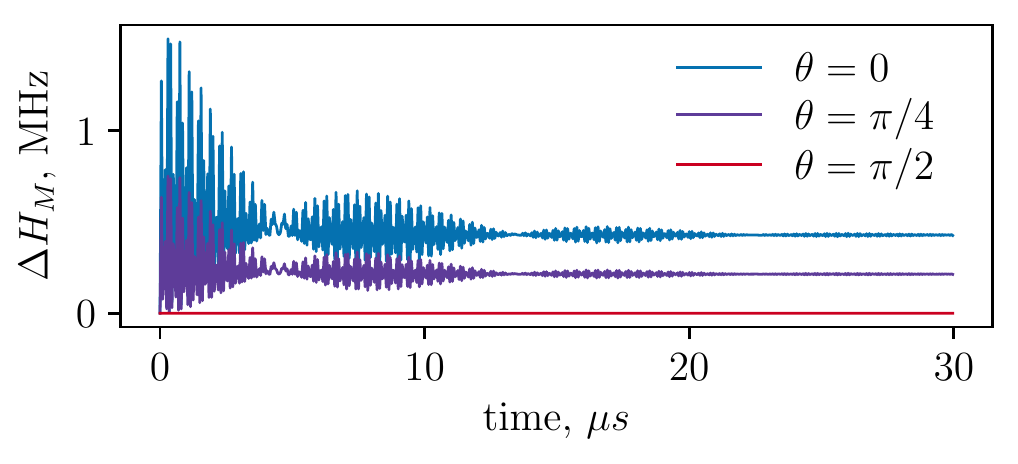}
\caption{\label{fig:EnergyTransfer} \new{The average energy of the membrane as a function of time for coherent states of the light field containing $6 \times 10^6$ photons. Perfectly distinguishable, partially distinguishable and perfectly indistinguishable gases are denoted by red, purple and blue lines. We utilise the following parameters from the experimental settings in \cite{experiment2,experiment5}: $\omega_M =350$kHz, $\omega =20$THz, $\lambda=34$GHz, $\kappa=$85kHz, $\kappa_M=$1Hz, $m=$45ng and $g x_{\mbox{\tiny zpf}}=$3.3kHz, with similar behaviour expected for a range of parameters.}}
\end{figure}

In the limit in which the cavity damping is much faster than the membrane damping, as is the case in experimental settings such as \cite{experiment2,experiment4}, the energy of the membrane, as shown in Fig.~\ref{fig:EnergyTransfer}, tends to an approximately constant value on the time scale $\frac{1}{\kappa} \ll t \ll \frac{1}{\kappa_M}$. In this limit the energy transfer to the membrane after being driven by a single pair of pulses is
\begin{equation}\label{eq:CoherentTrans}
      \Delta H_M^{t \gg \frac{1}{\kappa}} = \mu |\alpha|^2 + \eta  \left( |\alpha|^2 +  |\alpha|^4 \cos^2(\theta) \right) \; ,
\end{equation}
where $\eta = 1.2 \times 10^{-8}$Hz and $ \mu = 1.3 \times 10^{-18}$Hz for the experimental parameters listed in Fig.~\ref{fig:EnergyTransfer}.
For a pulse containing $6 \times 10^6$ indistinguishable photons, the expected energy transfer to the membrane is of the order of 400kHZ. This effect could be amplified by driving the cavity with a train of laser pulses, increasing the viability of experimentally observing the enhanced energy transfer to the membrane using currently available measurement protocols~\cite{CooledMembrane}}.

It is natural to ask whether this energy transfer to the piston membrane, $\Delta H_M$, should be interpreted as heat or work.
While the question of how to define work~\cite{defofquantumwork2,defofquantumwork3,defofquantumwork4, defofquantumwork3.5} and heat~\cite{CirilQuantumHeat, HeatFelix} in the quantum regime has been discussed extensively,
in essence the distinction reduces to the extent to which the energy is `useful' energy as opposed to un-directed fluctuating energy.
Since the quantum mechanical average of the mechanical displacement and momentum vanishes at all times, the energy transfer $\Delta H_M$ is entirely given in terms of the fluctuations resulting from the entanglement between light fields and mechanical degree of freedom~\cite{Zurek}.
In this vein, one might classify the energy transfer as heat rather than work.

However, the fact that the quantum mechanical average over displacement vanishes can be seen as a direct consequence of the system's mirror symmetry ({\it i.e.} exchange of $L_p$ and $R_p$ and simultaneous replacement of $X_M$ with $-X_M$). Given a symmetric initial state, this symmetry is preserved during the dynamics and necessarily needs to be satisfied in the final state. Nonetheless, this symmetry could be broken with a measurement of the photon number in the left or right part of the cavity. As indicated by the correlations depicted in Fig.~\ref{fig:g2} and Fig.~\ref{fig:HOMinCavity}, a suitable measurement \new{will collapse the symmetric superposition} and therefore is likely to find a pronounced misbalance of photons between left and right corresponding to a substantial \new{instantaneous} displacement of the membrane.
Indeed, the cross correlation function
\begin{equation}\label{eq:DeltaNXM}
\langle \Delta N(t) X_M(t) \rangle= \nu(t) \delta N +  \zeta(t)  \left( \langle N\rangle +  \langle N \rangle^2 \cos^2(\theta)\right) \, ,
\end{equation}
between the photon number difference, $\Delta N = N_L  - N_R$, and the displacement of the membrane,
with $\nu(t)$ and $\zeta(t)$ oscillatory prefactors depending only the system parameters~\cite{supMat}, features the same quadratic enhancement for indistinguishable photons as found for the energy transfer, Eq.~\eqref{eq:DH}.
This suggests that a reasonably simple Szilard-type extraction protocol~\cite{BunchingSzilardPRL}, using auxiliary measurements on the light field, would allow one to find a predictable displacement of the membrane that increases with the indistinguishability of the photons in the cavity. The potential energy associated with this displacement is well defined and thus could plausibly be interpreted as a work output.

\medskip

\new{The bunching enhanced energy transfer to the piston membrane for indistinguishable photons draws a link between iconic thermodynamic experiments conceived by Gibbs, Maxwell and Szilard, and a paradigmatic example of the impact of indistinguishability in quantum optics, the HOM effect.
The optomechanical analysis further gives a flavor of the rich physics that can be explored by explicitly introducing polarisation into optomechanical setups, while introducing a new platform for quantum thermodynamic experiments. For example,}
a crucial difference between the present optomechanical setting and classical thermodynamical experiment, is the inability of the photons to thermalize via mutual interactions.
Interactions with dye-molecules on the other hand are routinely used to mediate effective interactions between photons resulting in thermalization~\cite{klaers2010,tinyBEC}.
One may thus envision extensions of the presently discussed setup with thermalization rates as additional parameters, permitting a broad range of future directions. \new{Other open questions include the variation of initial state, optomechanical coupling regime and coupling of the photons to the heat bath.}
\new{Similarly to the present analysis, such settings} can be discussed as a thought experiment or even realized in practice with optomechanical systems.

\bigskip
\acknowledgements
\new{We are grateful for insightful conversations with Jack Clarke and Michael Vanner.} We acknowledge support from the Engineering and Physical Sciences Research Council Centre for Doctoral Training in Controlled Quantum Dynamics; the Engineering and Physical Sciences Research Council Grants EP/M009165/1 and EP/S000755/1 and the Royal Society.

\bibliography{refs}

\clearpage

\appendix

\onecolumngrid

\section{Supplementary Information for ``Enhanced energy transfer to an optomechanical piston from indistinguishable photons"}

Here we provide a derivation of the results presented in the main text.

\medskip

\medskip

\paragraph*{Initial condition}

We make the following assumptions on the initial state of the light field and the mechanical \change{oscillator}:
\begin{enumerate}
    \item The photons and membrane are initially non-interacting such that $\langle A(0) B(0)  \rangle = \langle A(0) \rangle \langle B(0)  \rangle \,$ for any operator of the light field, $A$, and operator of the membrane, $B$.
    \item  All photons in the left cavity are in the polarisation state $\ket{V}$ and all photons in the right part are in the polarization state $\ket{\theta}$. 
    \item The number distribution of the photons in the left part of the cavity is the same as in the right part but the precise form of this distribution can be freely chosen. That is, $\langle f(N_L) \rangle = \langle f(N_R) \rangle$ for all functions $f$. 
    \item The average initial position and momentum of membrane is zero, that is $\langle X_M(0) \rangle = \langle P_M(0) \rangle = 0 $ where $X_M$ and $P_M$ are the position and momentum operators of the mechanical degree of freedom. 
\end{enumerate}

\medskip

\paragraph*{Calculation of dynamics}

The membrane evolves as a quantum harmonic \change{oscillator} driven by the radiation pressure from the photons in the cavity. To explicitly calculate the dynamics of the membrane we work in the Heisenberg picture since this allows us to obtain results for the evolution of pertinent physical variables that hold for the general set of initial conditions stated above. Working in the Heisenberg picture, the equation of motion of the membrane reads 
\begin{equation}\label{Eq:MotionMembrane}
    \frac{d^2 X_M}{d t^2} + 2 \kappa_M \frac{d X_M}{d t} +  \omega_M^2 X_M = \frac{F(g)}{m} 
\end{equation}
where  the driving force $F(g) = g \Delta N$ and $\kappa_M$ is the damping rate of the membrane.

To calculate the force exerted by the photons on the membrane we need to determine the dynamics of the photons. The left and right modes (with polarisations $p=H$ and $p=V$) obey 
\begin{align}\label{eq:EquationOfMotionModes}
    \frac{d L_p}{dt} = -i (\omega + g X_M - i \kappa) L_p - i \frac{\lambda}{2} R_p \\
    \frac{d R_p}{dt} = -i (\omega - g X_M - i \kappa) R_p - i \frac{\lambda}{2} L_p \; ,
\end{align}
where we include phenomenologically the damping of the cavity modes at the rate $\kappa$. 

Solving these coupled differential equations, Eq.~\eqref{Eq:MotionMembrane} and Eq.~\eqref{eq:EquationOfMotionModes}, for the four photonic modes and single mechanical mode exactly is prohibitively difficult but a perturbative solution that describes the dynamics of selected observables for any given initial state can be constructed. Given the typically weak coupling between the optical and mechanical degrees of freedom, it is appropriate to consider the system dynamics perturbatively in the coupling $g$. To first order, which we focus on, this amounts to assuming that the back action of the motion of the membrane on the photon dynamics is negligible. 

To calculate the dynamics of the membrane to 1st order in $g$, we drop the dependence of Eq.~\eqref{eq:EquationOfMotionModes} on $g$ and solve the resulting coupled differential equations to find that 
\begin{align}\label{eq:RL}
        L_p(t) = \exp\left(-(\kappa + i\omega) t\right) \left( L_p (0) \cos \left(\frac{\lambda t}{2} \right) - i R_p (0) \sin \left(\frac{\lambda t}{2} \right) \right)  \\
        R_p(t) = \exp\left(-(\kappa + i\omega) t\right) \left( R_p (0) \cos \left(\frac{\lambda t}{2} \right) - i L_p (0) \sin \left(\frac{\lambda t}{2} \right) \right) 
\end{align}
From this it follows that the membrane is driven by the force
\begin{equation}\label{eq:ForceSol}
    F(g) = g \exp(- 2 \kappa t) \left( \Delta N(0) \cos(  \lambda t ) + \Delta K (0) \sin( \lambda t) \right) \; ,
\end{equation}
where we have defined $\Delta K := \Delta K_H + \Delta K_V$  with 
\begin{equation}
     \Delta K_p := i ( R_p^\dagger L_p -  R_p L_p^\dagger) \ \ \ \text{for} \ \  \ p= H, V \; . 
\end{equation}

The solution to the equation of motion of the membrane, Eq.~\eqref{eq:ForceSol}, for the time dependent force induced by the dynamics of the photons, Eq.~\eqref{eq:ForceSol}, is of the form 
\begin{equation}\label{eq:TrialSol}
    X_M(t) = c(t) \Delta N(0)  + d(t) \Delta K (0)  + h(t) X_M(0) + j(t) P_M(0)  \; ,
\end{equation}
where the functions $a(t), b(t), c(t)$ and $d(t)$ depend on the system parameters ($g, \omega_M, \lambda, m, \kappa$ and $\kappa_M$) and not on the initial state. The functions $a(t), b(t), c(t)$ and $d(t)$ can be calculated explicitly by substituting the trial solution Eq.~\eqref{eq:TrialSol} into Eq.~\eqref{Eq:MotionMembrane}; however, since the functions do not depend on the initial state of the photons it is not necessary to calculate these functions to determine the distinguishability dependence of the energy transfer to the membrane. 
The average energy of the membrane can be written in terms of the membrane displacement operator as 
\begin{equation}
    \langle H_M(t) \rangle = \left\langle  \frac{m \omega_M^2 X_M(t)^2}{2} + \frac{m  \dot{X}_M(t)^2}{2}  \right\rangle \; 
\end{equation}
and therefore 
\begin{equation}
\begin{aligned}
    \langle H_M(t) \rangle = \frac{m \omega_M^2 }{2} \left\langle (c(t) \Delta N(0)  + d(t) \Delta K (0)  + h(t) X_M(0) + j(t) P_M(0) )^2 \right\rangle \\ + \frac{m}{2} \left\langle (\dot{c}(t) \Delta N(0)  + \dot{d}(t) \Delta K (0)  + \dot{h}(t) X_M(0) + \dot{j}(t) P_M(0) )^2 \right\rangle \; .
\end{aligned}
\end{equation}
Since we assume the light and mechanical modes are initially uncorrelated and that the mean position and momentum of the membrane vanish, it follows that the initial correlation functions $\langle \Delta N(0) X_M(0) \rangle $, $\langle \Delta K(0) X_M(0) \rangle $,  $\langle \Delta N(0) P_M(0) \rangle $ and $\langle \Delta K(0) P_M(0) \rangle $ vanish. Similarly, since we assume that the photons in the left cavity are in polarisation state $\ket{V}$ and the photons in the right cavity are in the polarisation state $\ket{\theta}$, the $\langle \Delta K(0) \Delta N(0) \rangle $ and  $\langle \Delta K(0) \Delta N(0) \rangle $ terms also vanish. It follows that the energy of the membrane is given by 
\begin{equation}
    \langle H_M(t) \rangle =  \langle H_M(0) \rangle + \frac{1}{2}\left( m \omega_M^2 c(t)^2  + m\dot{c}(t)^2 \right) \left\langle \Delta N(0)^2  \right\rangle + \frac{1}{2}\left( m \omega_M^2 d(t)^2  + m\dot{d}(t)^2 \right) \left\langle \Delta K(0)^2  \right\rangle \ ,
\end{equation}
and therefore the time dependent energy transfer to the membrane, $\Delta H_M(t) = \langle H_M(t) \rangle - \langle H_M(0) \rangle$, is 
\begin{equation}
 \Delta H_M(t) =  \frac{u(t)}{2} \left\langle \Delta N(0)^2  \right\rangle + \frac{v(t)}{2} \left\langle \Delta K(0)^2  \right\rangle \ ,
\end{equation}
where $u(t) = m \omega_M^2 c(t)^2  + m\dot{c}(t)^2 $ and $v(t) = m \omega_M^2 d(t)^2  + m\dot{d}(t)^2 $. 

\medskip

The distinguishability and photon number dependence of the energy of the membrane enter through the evaluation of the $\langle \Delta N(0)^2 \rangle$ and $\langle \Delta K(0)^2 \rangle$ terms. The $\langle \Delta N(0)^2 \rangle$ term is calculated as follows 
\begin{equation}
\begin{aligned}
    \langle \Delta N(0)^2 \rangle &=  \langle (N_L(0) - N_R(0))^2 \rangle \\ &=  \langle N_L(0)^2 + N_R(0)^2 - 2 N_L(0) N_R(0) \rangle \\
    &= 2 \langle N^2 \rangle - 2 \langle N \rangle^2
\end{aligned}
\end{equation}
where the final line follows from the fact that we assume the photon number distribution in the two halves of the cavity are equal \change{and uncorrelated} and therefore we can write $\langle N \rangle := \langle N_L(0) \rangle = \langle N_R(0) \rangle$ and $\langle N^2 \rangle := \langle N_L(0)^2 \rangle = \langle N_R(0)^2 \rangle $. Thus the $\langle \Delta N(0)^2 \rangle$ is distinguishability independent and equal to the variance, $\delta N$, in the number of photons in each gas,
\begin{equation}
    \langle \Delta N(0)^2 \rangle  =  \langle N^2 \rangle -  \langle N \rangle^2 := \delta N\; .
\end{equation} 
Similarly, for the $\langle \Delta K(0)^2 \rangle$ term we have that 
\begin{equation}
\begin{aligned}
\langle \Delta K(0)^2 \rangle& = \langle  (\Delta K_H(0) + \Delta  K_V(0))^2 \rangle \\ &= \langle  \Delta K_H(0)^2 + \Delta  K_V(0))^2 + 2  K_H(0)  K_V(0)\rangle \ .
\end{aligned}
\end{equation}
The cross term vanishes since all the photons in the left cavity are vertically polarised and remaining two terms evaluate to 
\begin{equation}
\begin{aligned}
\langle \Delta K(0)^2 \rangle &= \langle  \Delta K_H(0)^2 + \Delta  K_V(0))^2 \rangle \\ &= \langle L_H^\dagger L_H  \rangle + \langle  R_H^\dagger R_H  \rangle + 2 \langle L_H^\dagger L_H \rangle \langle  R_H^\dagger R_H  \rangle + \langle L_V^\dagger L_V  \rangle + \langle  R_V^\dagger R_V  \rangle + 2 \langle L_V^\dagger L_V \rangle \langle  R_V^\dagger R_V  \rangle \\  &= 2 \langle N \rangle + 2 \langle N \rangle^2 \cos^2(\theta)\ . 
\end{aligned}
\end{equation}
Substituting these expressions into the equation for the total energy of the membrane we have that the change in energy of the membrane is given by
\begin{equation}\label{eq:DistDepend}
       \Delta H_M(t)  =  u(t) \delta N +  v(t)  \left( \langle N \rangle +  \langle N \rangle^2 \cos^2(\theta)\right) \; .
\end{equation}
We thus see that the energy transfer to the membrane is quadratically enhanced for indistinguishable photon gases. The derivation of the membrane correlation function $\langle \Delta N(t) X_M(t) \rangle$, Eq.~\eqref{eq:DeltaNXM}, proceeds in an entirely analogous manner to the derivation of $\Delta H_M(t)$ set out here. 

\medskip

In the general case the expressions for $u(t)$ and $v(t)$ are long and uninstructive. However, in the experimentally realisable limit in which the cavity damping $\kappa$ is much faster than the membrane damping $\kappa_M$, the energy of the membrane tends to an approximately constant value on the time scale $\frac{1}{\kappa} \ll t \ll \frac{1}{\kappa_M}$ as shown in Fig.~\ref{fig:EnergyTransfer}. In this limit we can effectively disregard the membrane damping, that is set $\kappa_M = 0$, and calculate the time averaged change in energy of the membrane in the limit of large $t$, 
\begin{equation}
      \Delta H_M^{t \gg \frac{1}{\kappa}} = \lim_{ \substack{\kappa_M \rightarrow 0 \\ t \rightarrow \infty}} \langle H_M(t) - H_M(0) \rangle  \, . 
\end{equation}
To perform this calculation we disregard any terms that exponentially vanish in the limit of large $t$ and average over remaining sinusoidal terms. On doing so we obtain,
\begin{equation}
      \Delta H_M^{t \gg \frac{1}{\kappa}} =  \mu \delta N + \eta  \left( \langle N \rangle +  \langle N \rangle^2 \cos^2(\theta)\right) 
\end{equation}
where 
\begin{align}
        &\mu = \frac{2g^2}{m}  \frac{2(4\kappa^2 + \omega_M^2) + (\lambda^2 + \omega_M^2) \delta_{\kappa,0}}{(4 \kappa^2 + (\lambda - \omega_M)^2) (4 \kappa^2 + (\lambda + \omega_M)^2) } \\ 
        &\eta = \frac{g^2}{m}  \frac{2 \lambda^2 + (\lambda^2 + \omega_M^2)\delta_{\kappa,0} }{(4 \kappa^2 + (\lambda - \omega_M)^2) (4 \kappa^2 + (\lambda + \omega_M)^2)} \, . 
\end{align}
In the idealised case of a friction-less piston and no cavity damping, that is for $\kappa = 0$ and $\kappa_M = 0$, these expressions reduce to 
\begin{align}
        &\mu = \frac{2g^2}{m}  \frac{\lambda^2 + 3 \omega_M^2 }{ (\lambda^2 + \omega_M^2)^2} \\ 
        &\eta = \frac{g^2}{m}  \frac{ 3\lambda^2 + \omega_M^2}{(\lambda^2 + \omega_M^2)^2)} \, ,
\end{align}
with the ratio of the prefactors given by 
\begin{equation}
    \frac{\eta}{\mu} = \frac{ 3\lambda^2  + \omega_M^2 }{2(\lambda^2 + 3 \omega_M^2)}  \; .
\end{equation}
Therefore the relative significance of the distinguishability dependent contribution to the energy transfer term depends on the relative magnitude of the tunnelling rate $\lambda$ and the membrane frequency $\omega_M$. However, in the presence of damping, that is for $\kappa \neq 0$, these expressions reduce to
\begin{align}
        &\mu = \frac{2g^2}{m}  \frac{2(4\kappa^2 + \omega_M^2)}{(4 \kappa^2 + (\lambda - \omega_M)^2) (4 \kappa^2 + (\lambda + \omega_M)^2)} \\ 
        &\eta = \frac{2g^2}{m}  \frac{ \lambda^2}{(4 \kappa^2 + (\lambda - \omega_M)^2) (4 \kappa^2 + (\lambda + \omega_M)^2)} \, ,
\end{align}
with the ratio of the prefactors given by 
\begin{equation}
    \frac{\eta}{\mu} = \frac{ \lambda^2}{2(4\kappa^2 + \omega_M^2)}  \; .
\end{equation}
When $\lambda$ is substantially greater than $\kappa$ and $\omega_M$, 
as is the case in the experimental settings such as \cite{experiment2,experiment4}, $\eta$ is substantially larger than $\mu$ and therefore the distinguishability dependent term in Eq.~\ref{eq:DistDepend} dominates over the fluctuation dependent term.

\medskip

The effect of the back action of the membrane's motion on the photons dynamics can be incorporated by calculating the evolution of the membrane's displacement to second order in $g$. As shown in Fig.~\ref{fig:SecondOrderEnergyTransfer} below, the second order contribution to the energy transfer to the membrane also increases with the indistinguishability of the photons. In contrast to the first order contribution, the second order contribution depends on the initial state of the membrane. Assuming the membrane is initially thermal, the second order contribution increases with the initial temperature of the membrane as shown in Fig.~\ref{fig:SecondOrderEnergyTransfer}. Also, in contrast to the first order, the second order contribution to the energy of the membrane, once averaged over oscillations, grows quadratically over time. Nonetheless, the effect of back action is negligible over the lifetime of the cavity, $1/\kappa$, as long as the initial thermal occupancy of the membrane, $\bar{n}_{th}$, is lower than $\sim50$ phonons. Such low temperature regimes have been experimentally achieved~\cite{CooledMembrane}.

\begin{figure}[h!]
\centering
\includegraphics[width=0.8\linewidth]{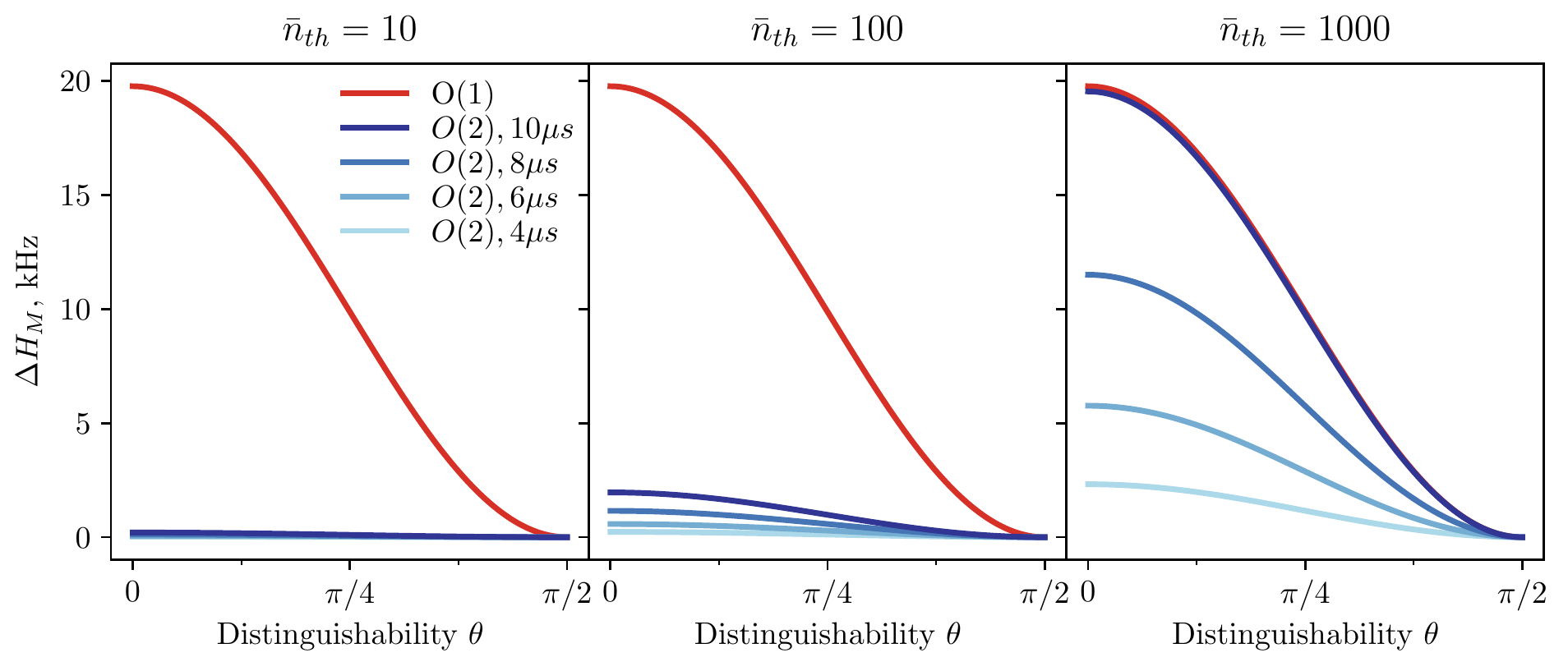}
\caption{\label{fig:SecondOrderEnergyTransfer} We plot the first and second order contributions to the time averaged energy transfer to the membrane as a function of the distinguishability of the photons in the cavity in the absence of damping effects ($\kappa = \kappa_C = 0$). Once the energy transfer to the membrane is averaged over fast oscillations the first order contribution to the energy transfer is constant in time, while the second order contribution has an amplitude that grows quadratically in time. Here we plot the second order correction up to $10\mu s$ (the cavity lifetime in the experimental setup of Ref[23]). The membrane is prepared in a thermal state containing on average $\bar{n}_{th}$ phonons, where $\bar{n}_{th} = 10$, $\bar{n}_{th} = 100$ and $\bar{n}_{th} = 1000$ in the left, centre and right plots respectively and the light field in both halves of the cavity is initially in a coherent state containing on average $10^6$ photons.  We utilise the following parameters from the experimental settings in \cite{experiment2,experiment5}: $\omega_M =350$kHz, $\omega =20$THz, $\lambda=34$GHz, $m=45$ng and $g x_{\mbox{\tiny zpf}}=$3.3kHz, with similar behaviour expected for a range of parameters.}
\end{figure}  


\medskip

\paragraph*{Photon correlation function calculation}

The degree of bunching of the photons in the cavity is quantifiable by the second order coherence correlation function $\mathbb{g}$. The second-order correlation function between any two modes $a$ and $b$ is given by
\begin{equation}
   \mathbb{g}_{a,b}(t) := \frac{\langle a^\dagger(t) b^\dagger(t) b(t) a(t)\rangle}{N_a(t) N_b(t)} \; ,
\end{equation}
where $\langle a^\dagger(t) b^\dagger(t) b(t) a(t)\rangle$ is the probability of measuring a photon in mode $a$ and a photon in mode $b$ at time $t$, and the product of 
\begin{equation}
\begin{aligned}
&N_a(t) = \langle a^\dagger(t) a(t)\rangle \ \ \ \text{and} \ \ \ N_b(t) =  \langle b^\dagger(t)b(t)\rangle \  
\end{aligned}
\end{equation}
is a normalisation factor. 

The second order correlation function is most often used to quantify the spacing of photons within a \textit{single} mode, i.e. where $a = b$. A coherent light beam has randomly spaced photons and therefore a $\mathbb{g}_{a,a}$ value of 1. If $\mathbb{g}_{a,a}$ is greater than 1 the probability of measuring two photons simultaneously is greater than expected for a random beam of photons and in this sense the photons are `bunched' together. Thermal light with super-Poissonian statistics and a $\mathbb{g}_{a,a}$ value of 2 is therefore bunched. In contrast, if $\mathbb{g}_{a,a}$ is less than 1 then the probability of measuring two photons simultaneously is greater than expected for a random beam of photons and the photons are said to be `antibunched'. The correlation function corresponding to an $n$ photon Fock state scales as $1- \frac{1}{n}$.
Therefore a single Fock state photon is maximally antibunched with a $\mathbb{g}_{a,a}$ of 0 but the degree of antibunching decreases as the number of photons is increased~\cite{Fox, Loudon}. 

We are primarily interested in the correlations between the \textit{between} the left and right modes, rather than within a mode, that is $\mathbb{g}_{\mbox{\tiny L,R}}(t)$ which can be written as
\begin{equation}
\mathbb{g}_{\mbox{\tiny L,R}}(t)= \frac{\langle N_L(t)N_R(t)\rangle}{\langle N_L(t)\rangle  \langle N_R(t)\rangle}\ .
\end{equation}
A $\mathbb{g}_{L,R}$ value of 0 entails that if we simultaneously measure the number of photons in the left and right halves of the cavity, we are certain to find at least one photon in the left cavity and zero photons in the right cavity, or vice versa, despite on average there being photons in both halves of the cavity. Thus, in contrast to the single mode correlation function, a $ \mathbb{g}_{\mbox{\tiny L,R}}$ value of 0 indicates that the correlations between the photons in the two halves of the cavity are such that the photons are bunched in a single cavity. Conversely, a $ \mathbb{g}_{\mbox{\tiny L,R}}$ of 1 indicates that whenever a photon is measured in the one cavity, it is certain that at least one photon will be measured in the other cavity, indicating that the photons are evenly spread between the two halves of the cavity. In this way, we can use $ \mathbb{g}_{\mbox{\tiny L,R}}$ to quantify the degree to which on measurement the photons bunch in one half of the optical cavity. 

\medskip


To verify that bunching is playing an active role in the dynamics we calculate the correlation function $\mathbb{g}_{\mbox{\tiny L,R}}(t)$ for cavities containing a BS membrane. To do so, we use the expressions for the evolution of the creation and annihilation operators of the cavity modes on disregarding the back-action of the membrane on the photon dynamics, Eq.~\eqref{eq:RL}. We find that the number operators for the photons in the left and right halves of the cavity evolve as 
\begin{align}
    &N_L(t) = \cos^2\left(\frac{\lambda t}{2}\right) N_L(0)  + \sin^2\left(\frac{\lambda t}{2}\right) N_R(0)  + \frac{\sin(\lambda t)}{2} \Delta K   \\
    &N_R(t) = \cos^2\left(\frac{\lambda t}{2}\right) N_R(0)  + \sin^2\left(\frac{\lambda t}{2}\right) N_L(0)  - \frac{\sin(\lambda t)}{2} \Delta K 
\end{align}
and therefore the correlation function evaluates to 
\begin{equation}
\begin{aligned}\label{eq:g2eval}
\mathbb{g}_{\mbox{\tiny L,R}}(t) &= \frac{1}{\langle N_L(0) \rangle \langle N_R(0) \rangle } \left( \frac{\cos(\lambda t)+3}{4}\langle N_L(0) \rangle \langle N_R(0) \rangle + \frac{\sin(\lambda t)^2}{4} (\langle N_L(0)^2\rangle + \langle N_R(0)^2\rangle - \langle \Delta K^2 \rangle ) \right) \\
&= \frac{1}{4 \langle N \rangle^2} \left( (\cos(\lambda t)+3) \langle N \rangle^2 + 2\sin^2(\lambda t)(\langle N^2 \rangle - \langle N \rangle - \langle N \rangle^2 \cos^2(\theta) ) \right) \\
&= \frac{1}{4} \left( (\cos(\lambda t)+3) + 2\sin^2(\lambda t)(\mathbb{g}_{\mbox{\tiny L,L}}(0) - \cos^2(\theta) ) \right) \; ,
\end{aligned}
\end{equation}
where $\mathbb{g}_{\mbox{\tiny L,L}}(0) = \mathbb{g}_{\mbox{\tiny R,R}}(0) = \frac{\langle N(N-1) \rangle}{\langle N \rangle^2}$ is the second order correlation function of both the photon gases. 

To highlight the distinguishability dependence of the correlation function it is insightful to consider the quantity $1 - \langle  \mathbb{g}_{\mbox{\tiny L,R}}(t) \rangle_t$ where $\langle \mathbb{g}_{\mbox{\tiny L,R}}(t) \rangle_t$ denotes the time average of $ \mathbb{g}_{\mbox{\tiny L,R}}(t)$. 
This quantity is a measure of the degree of bunching in the cavity. More precisely, it equals 0 when the photons are evenly spread between the two halves of the cavity but is positive for light bunched in a single cavity and negative for light that is antibunched between the two parts. Using Eq.~\ref{eq:g2eval}, we thus find that 
\begin{equation}\label{eq:BunchingBS}
    1 - \langle  \mathbb{g}_{\mbox{\tiny L,R}}(t) \rangle_t = \frac{1}{4} \left(1 + \cos^2(\theta) - \mathbb{g}_{\mbox{\tiny L,L}}(0) \right) \, .
\end{equation}
Therefore,  for all possible photon number distributions of the initial gases, the bunching within one half of the cavity decreases with the distinguishability of the photons.



\end{document}